\documentclass[12pt]{article}
\usepackage{amssymb,latexsym}
\usepackage[dvips]{graphicx}
\newtheorem{theo}{Theorem}
\newtheorem{defi}[theo]{Definition}

\def\bea{\begin{eqnarray}}
\def\eea{\end{eqnarray}}
\def\nn{\nonumber}
\def\beq{\begin{equation}}
\def\eeq{\end{equation}}

\def\nn{\nonumber}

\def\Z{\mathbb{Z}}
\def\ra{\rangle}
\def\la{\langle}

\renewcommand{\theequation}{\arabic{section}.\arabic{equation}}
\begin{document}
\begin{center}
{\Large \bf A classification of generalized quantum statistics}\\[4mm]
{\Large \bf associated with classical Lie algebras}\\[3cm]
{\bf N.I.\ Stoilova~\footnote{Permanent address:
Institute for Nuclear Research and Nuclear Energy, Boul.\ Tsarigradsko Chaussee 72,
1784 Sofia, Bulgaria} and J.\ Van der Jeugt}\\[2mm]
Department of Applied Mathematics and Computer Science,\\
University of Ghent, Krijgslaan 281-S9, B-9000 Gent, Belgium.\\
E-mails~: Neli.Stoilova@UGent.be, Joris.VanderJeugt@UGent.be.
\end{center}


\begin{abstract}
Generalized quantum statistics such as para-Fermi statistics is characterized
by certain triple relations which, in the case of para-Fermi statistics, are
related to the orthogonal Lie algebra $B_n=so(2n+1)$.
In this paper, we give a quite general definition of ``a generalized quantum
statistics associated to a classical Lie algebra $G$''.
This definition is closely related to a certain $\Z$-grading of $G$. The
generalized quantum statistics is then determined by a set of root vectors
(the creation and annihilation operators of the statistics) and the set
of algebraic relations for these operators.
Then we give a complete classification of all generalized quantum statistics
associated to the classical Lie algebras $A_n$, $B_n$, $C_n$ and $D_n$. 
In the classification, several new classes of generalized quantum statistics
are described.
\end{abstract}

\vspace{1cm}

\newpage
\renewcommand{\thesection}{\Roman{section}}
\renewcommand{\theequation}{\arabic{section}.{\arabic{equation}}}

\setcounter{equation}{0}
\section{Introduction} \label{sec:Introduction}%

In classical quantum statistics one works exclusively with Bose and Fermi 
statistics (bosons and fermions). A historically important extension or 
generalization of these quantum statistics has been known for 50 years, 
namely the para-Bose and para-Fermi statistics as developed by Green~\cite{Green}. 
Instead of the classical bilinear commutators or anti-commutators as for 
bosons and fermions, para-statistics is described by means of certain 
trilinear or triple relations. For example, for $n$ pairs of para-Fermi creation 
and annihilation operators $f_i^\xi$ ($\xi=\pm$ and $i=1,\ldots,n$), 
the defining relations are:
\begin{eqnarray}
&& [[f_{ j}^{\xi}, f_{ k}^{\eta}], f_{l}^{\epsilon}]=\frac 1 2
(\epsilon -\eta)^2
\delta_{kl} f_{j}^{\xi} -\frac 1 2  (\epsilon -\xi)^2
\delta_{jl}f_{k}^{\eta}, \label{para-Fermi} \\
&& \qquad\qquad \xi, \eta, \epsilon =\pm\hbox{ or }\pm 1;\quad j,k,l=1,\ldots,n. \nn 
\end{eqnarray}
About ten years after the introduction of para-Fermi relations by Green, 
it was proved that these relations are associated with the orthogonal 
Lie algebra $so(2n+1)=B_n$~\cite{KR}. More precisely, the Lie algebra generated 
by the $2n$ elements $f^\xi_i$, with $\xi=\pm$ and $i=1,\ldots,n$, 
subject to the relations~(\ref{para-Fermi}), is $so(2n+1)$ (as a Lie 
algebra defined by means of generators and relations). In fact, this 
can be considered as an alternative definition instead of the common 
definition by means of Chevalley generators and their known relations 
expressed by means of the Cartan matrix elements (inclusive the Serre relations). 
Moreover, there is a certain representation of $so(2n+1)$, the so-called 
Fermi representation ${\cal F}$, that yields the classical Fermi relations. 
In other words, the representatives ${\cal F}(f^\xi_i)$ satisfy the bilinear relations 
of classical Fermi statistics. Thus the usual Fermi statistics corresponds 
to a particular realization of para-Fermi statistics. 
For general para-Fermi statistics, a class of finite dimensional $so(2n+1)$ 
representations (of Fock type) needs  to be investigated.

Twenty years after the connection between para-Fermi statistics and the 
Lie algebra $so(2n+1)$, a new connection, between para-Bose statistics 
and the orthosymplectic Lie superalgebra $osp(1|2n)=B(0,n)$~\cite{Kac} 
was discovered~\cite{Ganchev}. 
The situation here is similar: the Lie superalgebra generated by $2n$ odd 
elements $b^\xi_i$, with $\xi=\pm$ and $i=1,\ldots,n$, subject to the triple
relations of para-Bose statistics, is $osp(1|2n)$ (as a Lie superalgebra 
defined by means of generators and relations). Also here there is a particular 
representation of $osp(1|2n)$, the so-called Bose representation ${\cal B}$, 
that yields the classical Bose relations, i.e. where the representatives ${\cal B}(b^\xi_i)$ 
satisfy the relations of classical Bose statistics. For more general 
para-Bose statistics, a class of infinite dimensional $osp(1|2n)$ 
representations needs to be investigated, and one of these representations 
corresponds with ordinary Bose statistics.

{}From these historical examples it is clear that para-statistics, 
as introduced by Green~\cite{Green} and further developed by many other 
research teams (see~\cite{Ohnuki} and  the references therein), 
can be associated with representations of the Lie (super)algebras of class $B$ 
(namely $B_n$ and $B(0,n)$). The question that arises is whether alternative
interesting types of generalized quantum statistics can be found in the framework of 
other classes of simple Lie algebras or superalgebras. In this paper we shall
classify all the classes of generalized quantum statistics for the classical 
Lie algebras $A_n$, $B_n$, $C_n$ and $D_n$, by means of their algebraic relations.
In a forthcoming paper we hope to perform a similar classification for the
classical Lie superalgebras.

We should mention that certain generalizations related to other Lie algebras 
have already been considered~\cite{Palev1}-\cite{Palev5}, 
although a complete classification was never made.
For example, for the Lie algebra $sl(n+1)=A_n$~\cite{Palev2}, a set of creation and annihilation
operators has been described, and it was shown that $n$ pairs of operators
$a^\xi_i$, with $\xi=\pm$ and $i=1,\ldots,n$, subject to the defining relations
\begin{eqnarray}
&& [[ a^+_i,a^-_j],a^+_k]= \delta_{jk} a^+_i + \delta_{ij} a^+_k, \nn\\
&& [[ a^+_i,a^-_j],a^-_k]= -\delta_{ik} a^-_j - \delta_{ij} a^-_k, \label{A-relations}\\
&& [a^+_i,a^+_j]=[a^-_i,a^-_j]=0, \nn
\end{eqnarray}
($i,j,k=1,\ldots,n$), generate the special linear Lie algebra $sl(n+1)$ 
(as a Lie algebra defined by means of generators and relations). 
Just as in the case of para-Fermi relations, (\ref{A-relations}) has two
interpretations. On the one hand, (\ref{A-relations}) describes the algebraic
relations of a new kind of generalized statistics, in this case $A$-statistics or
statistics related to the Lie algebra $A_n$. On the other hand, (\ref{A-relations})
yields a set of defining relations for the Lie algebra $A_n$ in terms of
generators and relations. Observe that certain microscopic and macroscopic 
properties of $A$-statistics have already been studied~\cite{PalevJeugt}-\cite{Jellal}.

The description (\ref{A-relations}) was given for the first time by 
N.\ Jacobson~\cite{Jacobson} in the context of ``Lie triple systems''. 
Therefore, this type of generators is often referred to as the 
``Jacobson generators'' of $sl(n+1)$. In this context, we shall mainly
use the terminology ``creation and annihilation operators (CAOs) for $sl(n+1)$''.

In the following section we shall give a precise definition of ``generalized
quantum statistics associated with a Lie algebra $G$'' and the corresponding
creation and annihilation operators. It will be clear that this notion
is closely related to gradings of $G$, and to regular subalgebras of $G$.
Following the definition, we go on to describe the actual classification
method. In the remaining sections of this paper, the classification results
are presented. The paper ends with some closing remarks and further outlook.

\setcounter{equation}{0}
\section{Definition and classification method} \label{sec:method}%

Let $G$ be a (classical) Lie algebra. A generalized quantum statistics associated with
$G$ is determined by a set of $N$ creation operators $x_i^+$ and $N$ annihilation
operators $x_i^-$. Inspired by the para-Fermi case and the example of $A$-statistics,
these $2N$ operators should satisfy certain conditions. First of all, these 
$2N$ operators should generate the Lie algebra $G$, subject to certain triple
relations like~(\ref{para-Fermi}) or~(\ref{A-relations}). Let $G_{+1}$ and $G_{-1}$
be the subspaces of $G$ spanned by these elements:
\begin{equation}
G_{+1} = \hbox{span} \{x^+_i;\ i=1\ldots,N\},\qquad
G_{-1} = \hbox{span} \{x^-_i;\ i=1\ldots,N\}.
\end{equation}
Then $[G_{+1},G_{+1}]$ can be zero (in which case the creation operators
mutually commute, as in~(\ref{A-relations})) or non-zero (as in~(\ref{para-Fermi})).
A similar statement holds for the annihilation operators and $[G_{-1},G_{-1}]$.
The fact that the defining relations should be triple relations, implies
that it is natural to make the following requirements:
\begin{eqnarray*}
&&[[x^+_i,x^+_j],x^+_k] = 0,\\
&&[[x^+_i,x^+_j],x^-_k] = \hbox{a lineair combination of } x^+_l,\\
&&[[x^+_i,x^-_j],x^+_k] = \hbox{a lineair combination of } x^+_l,\\
&&[[x^+_i,x^-_j],x^-_k] = \hbox{a lineair combination of } x^-_l,\\
&&[[x^-_i,x^-_j],x^+_k] = \hbox{a lineair combination of } x^-_l,\\
&&[[x^-_i,x^-_j],x^-_k] = 0.
\end{eqnarray*}
So let $G_{\pm 2}=[G_{\pm 1},G_{\pm 1}]$ and $G_0=[G_{+1},G_{-1}]$, 
then we may require $G_{-2} \oplus G_{-1} \oplus G_0 \oplus G_{+1}
\oplus G_{+2}$ (direct sum as vector spaces) to be a $\Z$-grading
of a subalgebra of $G$. Furthermore, since we want $G$ to be generated
by the $2N$ elements subject to the triple relations, one must have
$G= G_{-2} \oplus G_{-1} \oplus G_0 \oplus G_{+1} \oplus G_{+2}$.

There are two additional assumptions, again inspired by the known examples
(\ref{para-Fermi}) and (\ref{A-relations}). One is related to the fact
that creation and annihilation operators are usually considered to be 
each others conjugate. So, let $\omega$ be the standard antilinear anti-involutive
mapping of the Lie algebra $G$ (characterized by $\omega(x)=x^\dagger$
in the standard defining representation of $G$, 
where $x^\dagger$ denotes the transpose complex conjugate of the matrix
$x$ in this representation) then we should have $\omega(x_i^+)=x_i^-$.
And finally, we shall assume that the generating elements $x_i^\pm$
are certain root vectors of the Lie algebra $G$.

\begin{defi}
Let $G$ be a classical Lie algebra, with antilinear anti-involutive mapping $\omega$.
A set of $2N$ root vectors $x^\pm_i$ ($i=1,\ldots,N$) is called a set of
creation and annihilation operators for $G$ if:
\begin{itemize}
\item $\omega(x^\pm_i)=x^\mp_i$,
\item $G= G_{-2} \oplus G_{-1} \oplus G_0 \oplus G_{+1} \oplus G_{+2}$ is
a $\Z$-grading of $G$, with $G_{\pm 1}= \hbox{span}\{x^\pm_i,\ i=1\ldots,N\}$
and $G_{j+k}=[G_j,G_k]$.
\end{itemize}
The algebraic relations ${\cal R}$ satisfied by the operators $x_i^\pm$
are the relations of a generalized quantum statistics (GQS) associated with $G$.
\end{defi}

So a GQS is characterized by a set $\{x_i^\pm\}$ of CAOs
and the set of algebraic relations ${\cal R}$ they satisfy.
A consequence of this definition is that $G$ is generated by $G_{-1}$ and $G_{+1}$,
i.e.\ by the set of CAOs. Furthermore, since $G_{j+k}=[G_j,G_k]$, it follows that
\begin{equation}
G=\hbox{span}\{ x_i^\xi,\ [x_i^\xi,x_j^\eta]; \quad i,j=1,\ldots,N,\ \xi,\eta=\pm\}.
\end{equation} 
This implies that it is necessary and sufficient to give all relations of the following type:
\begin{itemize}
\item[(R1)] The set of all linear relations between the elements $[x_i^\xi, x_j^\eta]$ 
($\xi,\eta=\pm$, $i,j=1,\ldots,N$). 
\item[(R2)] The set of all triple relations of the form $[[x_i^\xi, x_j^\eta],x_k^\zeta]=
\hbox{linear combination of }x_l^\theta$. 
\end{itemize}
So in general ${\cal R}$ consists of a set of quadratic relations (linear combinations
of elements of the type $[x_i^\xi, x_j^\eta]$) and a set of triple relations.
This also implies that, as a Lie algebra defined by
generators and relations, $G$ is uniquely characterized by the set of generators
$x_i^\pm$ subject to the relations ${\cal R}$.

Another consequence of this definition is that $G_0$ itself is a subalgebra of $G$
spanned by root vectors of $G$, i.e.\ $G_0$ is a regular subalgebra of $G$.
Even more: $G_0$ is a regular subalgebra containing the Cartan subalgebra $H$ of $G$.
And by the adjoint action, the remaining $G_i$'s are $G_0$-modules.
Thus the following technique can be used in order to obtain a complete classification
of all GQS associated with $G$:
\begin{enumerate}
\item
Determine all regular subalgebras $G_0$ of $G$. If not yet contained in $G_0$, replace $G_0$
by $G_0 + H$.
\item
For each regular subalgebra $G_0$, determine the decomposition of $G$
into simple $G_0$-modules $g_k$ ($k=1,2,\ldots$).
\item
Investigate whether there exists a $\Z$-grading of $G$ of the form
\begin{equation}
G=G_{-2} \oplus G_{-1} \oplus G_0 \oplus G_{+1} \oplus G_{+2}, 
\label{5grading}
\end{equation}
where
each $G_i$ is either directly a module $g_k$ or else a sum of
such modules $g_1\oplus g_2\oplus \cdots$, such that
$\omega (G_{+i})=G_{-i}$.
\end{enumerate}

The first stage in this technique is a known one: to find regular subalgebras
one can use the method of extended Dynkin diagrams~\cite{Dynkin}. 
The second stage is straightforward by means of Lie algebra representation
techniques.
The third stage requires most of the work: one must try out all possible
combinations of the $G_0$-modules $g_k$, and see whether it is possible to
obtain a grading of the type~(\ref{5grading}). In this process, if one of the 
simple $G_0$-modules $g_k$ is such that $\omega(g_k)=g_k$, then it follows
that this module should be part of $G_0$. In other words, such a case
reduces essentially to another case with a larger regular subalgebra.

In general, when the rank of the semi-simple regular subalgebra is equal or close to the rank of $G$,
the corresponding $\Z$-grading of $G$ is ``short'' in the sense that
$G_i=0$ for $|i|>1$ or $|i|>2$. When the rank of the regular subalgebra becomes smaller,
the corresponding $\Z$-grading of $G$ is ``long'', and $G_i\ne 0$ for 
$|i|>2$. Thus the analysis shows that it is usually sufficient to
consider maximal regular subalgebras (same rank), or almost maximal
regular subalgebras (rank of $G$ minus 1 or 2).

Note that in~\cite{Palev5} a definition of CAOs was already given. Our Definition 1 
is inspired by the definition in~\cite{Palev5}, however it is different in the 
sense that the grading 
conditions $G_{j+k}=[G_j, G_k]$ are new. It is thanks to these new conditions that we
are able to give a complete classification of CAOs and the corresponding GQS.

In the following sections we shall give a summary of the classification
process for the classical Lie algebras $A_n$, $B_n$, $C_n$ and $D_n$.
Note that, in order to identify a GQS associated
with $G$, it is sufficient to give only the set of CAOs, or alternatively, to give
the subspace $G_{-1}$ (then the $x_i^-$ are the root vectors of $G_{-1}$,
and $x_i^+=\omega(x_i^-)$ ). The set ${\cal R}$ then consist of 
all quadratic relations (i.e. the linear relations between the elements
$[x_i^\xi, x_j^\eta]$) and all triple relations, and all of these
relations follow from the known commutation relations in $G$.
Because, in principle, ${\cal R}$ can be determined from the set $\{x_i^\pm; i=1,\ldots,N\}$, we
will not always give it explicitly. In fact, when $N$ is large, the corresponding relations
can become rather numerous and long. Such examples of GQS would be too complicated
for applications in physics. For this reason, we shall give
${\cal R}$ explicitly only when $N$ is not too large, more precisely when
$N$ is either equal to the rank of $G$ or at most double the rank of $G$.

Finally, observe that two different sets of CAOs $\{x_i^\pm; i=1\ldots,N\}$ and
$\{y_i^\pm; i=1\ldots,N\}$ (same $N$) are said to be isomorphic if, for a certain
permutation $\tau$ of $\{1,2,\ldots,N\}$, the relations between the elements
$x_{\tau(i)}^\pm$ and $y_i^\pm$ are the same. In that case, the regular subalgebra
$G_0$ spanned by $\{[x_i^+,x_j^-]\}$ is isomorphic (as a Lie algebra) to the
regular subalgebra spanned by $\{[y_i^+,y_j^-]\}$.

\setcounter{equation}{0}
\section{The Lie algebra $A_n=sl(n+1)$} \label{sec:A}%

Let $G$ be the special linear Lie algebra $sl(n+1)$, consisting of traceless
$(n+1)\times (n+1)$ matrices. The Cartan subalgebra $H$ of $G$ is the
subspace of diagonal matrices. The root vectors of $G$ are known to be the
elements $e_{jk}$ ($j\ne k=1,\ldots,n+1$), where $e_{jk}$ is a matrix with
zeros everywhere except a 1 on the intersection of row~$j$ and column~$k$.
The corresponding root is $\epsilon_j-\epsilon_k$, in the usual basis.
The anti-involution is such that $\omega(e_{jk})=e_{kj}$.
The simple roots and the Dynkin diagram of $A_n$ are given in Table~1,
and so is the extended Dynkin diagram.

In order to find regular subalgebras of $G=A_n$, one should delete nodes from
the Dynkin diagram of $G$ or from its extended Dynkin diagram.
We shall start with the ordinary Dynkin diagram of $A_n$, and subsequently
consider the extended diagram. 

\smallskip \noindent {\bf Step 1.} 
Delete node $i$ from the Dynkin diagram. The corresponding diagram is 
the Dynkin diagram of $sl(i)\oplus sl(n-i+1)$, so $G_0=H+sl(i)\oplus sl(n-i+1)$. 
In this case, there are only two $G_0$ modules and we can put
\begin{equation}
G_{-1}=\hbox{span}\{ e_{kl};\ k=1,\ldots ,i,\ l=i+1,\ldots ,n+1\}, \quad
G_{+1}=\omega(G_{-1}). \label{A1G-1}
\end{equation}
Therefore $sl(n+1)$ has the following grading:
\begin{equation}
sl(n+1)=G_{-1}\oplus G_0\oplus G_{+1},
\end{equation}
and the number of creation and annihilation operators is $N=i(n-i+1)$. 
Note that the cases $i$ and $n+1-i$ are isomorphic. 

The most interesting cases are those with $i=1$ and $i=2$, 
for which we shall explicitly give the relations ${\cal R}$ between the CAOs.

For $i=1$, $N=n$, the rank of $A_n$. Putting
\begin{equation} 
a_j^-=e_{1,j+1}, \qquad a_j^+=e_{j+1,1},\ j=1,\ldots ,n, \label{A1CAO}
\end{equation}
(for $A_n$, the possible sets $\{x_i^\pm\}$ will be denoted $\{a_i^\pm\}$,
for $B_n$, they will be denoted $\{b_i^\pm\}$, etc.) 
the corresponding relations ${\cal R}$ read ($j,k,l=1,\ldots,n$):
\begin{eqnarray}
 &&[a_j^+,a_k^+]=[a_j^-,a_k^-]=0,\nn \\
 &&[[a_j^+,a_k^-],a_l^+]=\delta_{jk}a_l^++\delta_{kl}a_j^+, \label{A1R} \\
 &&[[a_j^+,a_k^-],a_l^-]=-\delta_{jk}a_l^--\delta_{jl}a_k^-.  \nn
\end{eqnarray}
These are the relations of $A$-statistics~\cite{Palev1}-\cite{Palev2},
\cite{Palev5}-\cite{Jellal} as considered in the Introduction.

For $i=2$, $N=2(n-1)$, let 
\begin{eqnarray}
&& a_{-j}^-=e_{1,j+2}, \quad a_{+j}^-=e_{2,j+2},\qquad j=1,\ldots ,n-1, \nn \\
&& a_{-j}^+=e_{j+2,1}, \quad a_{+j}^+=e_{j+2,2},\qquad j=1,\ldots ,n-1. \label{A2CAO}
\end{eqnarray}
Now the corresponding relations are ($\xi, \eta, \epsilon =\pm; \ j,k,l=1,\ldots,n-1$):
\begin{eqnarray}
&&[a_{\xi j}^+,a_{\eta k}^+]=[a_{\xi j}^-,a_{\eta k}^-]=0,\nn\\
&& [a_{\xi j}^+,a_{-\xi k}^-]=0, \qquad j\neq k,  \nn \\
&& [a_{-j}^+,a_{- k}^-]=[a_{+j}^+,a_{+k}^-], \qquad j\neq k, \nn \\
&& [a_{+j}^+,a_{- j}^-]= [a_{+k}^+,a_{-k}^-], \label{A2R}\\
&& [a_{-j}^+,a_{+j}^-]= [a_{-k}^+,a_{+k}^-], \nn \\
&&[[a_{\xi j}^+,a_{\eta k}^-],a_{\epsilon l}^+]=\delta_{\eta\epsilon}\delta_{jk}
 a_{\xi l}^++\delta_{\xi\eta}\delta_{kl}a_{\epsilon j}^+, \nn \\
&&[[a_{\xi j}^+,a_{\eta k}^-],a_{\epsilon l}^-]=-\delta_{\xi \epsilon}\delta_{jk}
 a_{\eta l}^--\delta_{\xi\eta}\delta_{jl}a_{\epsilon k}^-. \nn 
\end{eqnarray}
These relations are already more complicated than (\ref{A1R}). But they
are still defining relations for the Lie algebra $A_n$.

\smallskip \noindent {\bf Step 2.} 
Delete node $i$ and $j$ from the Dynkin diagram.
By the symmetry of the Dynkin diagram, it is sufficient to consider
$1\leq i \leq \lfloor \frac{n}{2} \rfloor$ and $i< j <n+1-i$. 
We have $G_0=H+sl(i)\oplus sl(j-i)\oplus sl(n+1-j)$. In this case, 
there are six simple $G_0$-modules. All the possible combinations 
of these modules give rise to gradings of the form
\[
sl(n+1)=G_{-2}\oplus G_{-1}\oplus G_0\oplus G_{+1}\oplus G_{+2}.
\]  
There are essentially three different ways in which these $G_0$-modules
can be combined. To characterize these three cases, it is sufficient
to give only $G_{-1}$:
\begin{eqnarray}
G_{-1}&=&\hbox{span}\{ e_{kl},e_{lp};\ k=1,\ldots ,i,\ l=i+1,\ldots ,j,\ p=j+1, \ldots, n+1\},\label{A21}\\
&& \hbox{with }N=(j-i)(n+1-j+i); \nn\\
G_{-1}&=&\hbox{span}\{ e_{kl},e_{pk};\ k=1,\ldots ,i,\ l=i+1,\ldots ,j,\ p=j+1, \ldots, n+1\}, \label{A22}\\
&& \hbox{with }N=i(n+1-i); \nn\\
G_{-1}&=&\hbox{span}\{ e_{kl},e_{lp};\ k=1,\ldots ,i,\ p=i+1,\ldots ,j,\ l=j+1, \ldots, n+1\}, \label{A23}\\
&& \hbox{with }N=j(n+1-j).\nn 
\end{eqnarray}
It turns out that the sets of CAOs corresponding to (\ref{A22}) and
(\ref{A23}) are isomorphic to (\ref{A21}), so it is sufficient to consider 
only (\ref{A21}). Each case of (\ref{A21}) with 
$1\leq i \leq \lfloor \frac{n}{2} \rfloor$ and $i< j <n+1-i$ gives rise to 
a distinct GQS. For reasons explained earlier, we shall give the
corresponding set of relations explicitly only for small $N$.
In this case, it is interesting to give ${\cal R}$ for $j-i=1$, 
because then the number of creation or annihilation operators is $N=n$.
One can label the CAOs as follows:
\begin{eqnarray}
&&a_k^-=e_{k,i+1}, \quad a_k^+=e_{i+1,k}, \qquad k=1,\ldots ,i;\nn\\ 
&&a_k^-=e_{i+1,k+1}, \quad a_k^+=e_{k+1,i+1}, \qquad k=i+1,\ldots ,n.\label{A}
\end{eqnarray}
Using 
\begin{equation}
\langle k\rangle= \left\{ \begin{array}{lll}
 {0} & \hbox{if} & k=1,\ldots ,i \\ 
 {1} & \hbox{if} & k=i+1,\ldots ,n
 \end{array}\right.
\label{twokinds} 
\end{equation}
the quadratic and triple relations read:
\begin{eqnarray}
 &&[a_k^+,a_l^+]=[a_k^-,a_l^-]=0, \qquad k,l=1,\ldots, i \ \hbox{or}\
 k,l =i+1, \ldots ,n,\nn\\
 && [a_k^-,a_l^+]=[a_k^+,a_l^-]=0, \qquad k=1,\ldots , i, \ l=i+1,\ldots, n,\label{A21R} \\
 &&[[a_k^+,a_l^-],a_m^+]=(-1)^{\la l\ra +\la m \ra}\delta_{kl}a_m^++ 
 (-1)^{\la l\ra +\la m \ra} \delta_{lm}a_k^+ ,\ 
 k,l=1,\ldots, i \ \hbox{or}\
 k,l =i+1, \ldots ,n, \nn\\
&&[[a_k^+,a_l^-],a_m^-]=-(-1)^{\la l\ra +\la m \ra}\delta_{kl}a_m^--(-1)^{\la l\ra 
+\la m \ra}\delta_{km}a_l^- ,\
k,l=1,\ldots, i \ \hbox{or}\
 k,l =i+1, \ldots ,n, \nn\\
&&[[a_k^{\xi},a_l^{\xi}],a_m^{-\xi}]=-\delta_{km}a_l^{\xi}
+\delta_{lm}a_k^{\xi}, \
k=1,\ldots , i, \ l=i+1,\ldots, n,\nn \\
&&[[a_k^{\xi},a_l^{\xi}],a_m^{\xi}]=0 ,\qquad (\xi=\pm;\ k,l,m=1,\ldots,n).\nn 
\end{eqnarray}
The existence of the set of CAOs~(\ref{A}) is pointed out in~\cite{Palev1} as a possible
example. The relations~(\ref{A21R}) with $n=2m$ and $i=m$ are the commutation relations 
of the so called causal A-statistics investigated in~\cite{Palev4}.

\smallskip\noindent {\bf Step 3.} 
If we delete 3 or more nodes from the Dynkin diagram, the resulting
$\Z$-gradings of $sl(n+1)$ are no longer of the form $sl(n+1)=G_{-2}\oplus 
G_{-1}\oplus G_0\oplus G_{+1}\oplus G_{+2}$, but there would be non-zero
$G_i$ with $|i|>2$, so these cases are not relevant for our classification.

\smallskip\noindent {\bf Step 4.} 
Next, we move on to the extended Dynkin diagram of $G$. 
If we delete node $i$ from the extended Dynkin diagram,
then remaining diagram is again of type $A_n$, so $G_0=G$,
and there are no CAOs.

\smallskip\noindent {\bf Step 5.} 
If we delete node $i$ and $j$ from the extended Dynkin diagram ($0\leq i<j\leq n+1$),
then $sl(n+1)=G_{-1}\oplus G_0\oplus G_{+1}$ with $G_0=H+sl(j-i)\oplus sl(n-j+i+1)$, and
\[
G_{-1}=\hbox{span}\{ e_{kl};\ k=i+1\ldots ,j,\ l\neq i+1,\ldots ,j\}.
\]
The number of annihilation operators is $N=(j-i)(n+1-j+i)$.
It is not difficult to see that all these cases are isomorphic to those of Step~1.
This can also be deduced from the symmetry of the Dynkin diagram.

\smallskip\noindent {\bf Step 6.} 
If we delete nodes $i$, $j$ and $k$ from the extended Dynkin diagram ($i<j<k$),
then the corresponding $\Z$-gradings are of the form
\[
sl(n+1)=G_{-2}\oplus G_{-1}\oplus G_0\oplus G_{+1}\oplus G_{+2}.
\]  
All the corresponding CAOs, however, are isomorphic to those of Step~2
(which can again be seen from the remaining Dynkin diagram).

\smallskip\noindent {\bf Step 7.} 
If we delete 4 or more nodes from the extended Dynkin diagram, the corresponding
$\Z$-grading of $sl(n+1)$ has no longer the required properties (i.e.\ there are
non-zero subspaces $G_i$ with $|i|>2$).

\setcounter{equation}{0}
\section{The Lie algebra $B_n=so(2n+1)$} \label{sec:B}%

$G=so(2n+1)$ is the subalgebra of $sl(2n+1)$ consisting of matrices of the form:
\begin{equation}
\left(\begin{array}{ccc} a&b&c  \\
d&-a^t&e\\
-e^t&-c^t&0
\end{array}\right),
\label{so(2n+1)}
\end{equation}
where $a$ is any $(n\times n)$-matrix, $b$ and $d$ are antisymmetric $(n\times n)$-matrices, and
$c$ and $e$ are $(n\times 1)$-matrices. The Cartan subalgebra $H$ of $G$ is again the
subspace of diagonal matrices. The root vectors and corresponding roots of $G$ are given by:
\begin{eqnarray*}
e_{jk}-e_{k+n,j+n} & \leftrightarrow &\epsilon_j -\epsilon_k, \qquad j\neq k=1,\ldots ,n,\\
e_{j,k+n}-e_{k,j+n} & \leftrightarrow &\epsilon_j +\epsilon_k, \qquad j<k=1,\ldots ,n,\\
e_{j+n,k}-e_{k+n,j} & \leftrightarrow &-\epsilon_j -\epsilon_k, \qquad j<k=1,\ldots ,n,\\
e_{j,2n+1}-e_{2n+1,j+n} & \leftrightarrow &\epsilon_j,  \qquad j=1,\ldots ,n,\\
e_{n+j,2n+1}-e_{2n+1,j} & \leftrightarrow & -\epsilon_j , \qquad j=1,\ldots ,n.
\end{eqnarray*}
The anti-involution is such that $\omega(e_{jk})=e_{kj}$.
The simple roots, the Dynkin diagram and the extended Dynkin diagram of $B_n$ are given in Table~1.
Just as for $A_n$, we now start the process of deleting nodes from the Dynkin diagram or
from the extended Dynkin diagram.

\smallskip \noindent {\bf Step 1.}
Delete node~1 from the Dynkin diagram. The remaining diagram is that of $B_{n-1}$, so
$G_0=H+so(2n-1)\equiv H+B_{n-1}$. There are two $G_0$-modules:
\begin{equation}
G_{-1}=\hbox{span}\{ e_{1, 2n+1}-e_{2n+1,n+1},\ e_{1,k+n}-e_{k,n+1},\ 
 e_{1k}-e_{k+n,n+1};\ k=2,\ldots ,n\},\label{GB1}
\end{equation}
and $G_{+1}=\omega(G_{-1})$. Thus $so(2n+1)$ has the following grading:
\[
so(2n+1)=G_{-1}\oplus G_0\oplus G_{+1}
\]
and the number of (mutually commuting) creation and annihilation operators
is $N=2n-1$. Let us denote the CAOs by:
\begin{eqnarray}
&&b_{00}^-= e_{1, 2n+1}-e_{2n+1,n+1}, \ 
 b_{00}^+=e_{2n+1,1}-e_{n+1,2n+1}, \nn \\
&&b_{-k}^-= e_{1, n+k+1}-e_{k+1,n+1},\ b_{-k}^+= e_{n+k+1,1}-e_{n+1,k+1},\qquad k=1,\ldots ,n-1, \label{B1}\\
&&b_{+k}^-= e_{1, k+1}-e_{n+k+1,n+1},\ b_{+k}^+= e_{k+1,1}-e_{n+1,n+k+1},\qquad k=1,\ldots ,n-1. \nn
\end{eqnarray}
The  corresponding relations ${\cal R}$ are given by ($\xi,\eta,\epsilon=0,\pm;\ i,j,k=1,\ldots,n-1$):
\begin{eqnarray}
&&[b_{\xi i}^+, b_{\eta j}^+]= [ b_{\xi i}^-, b_{\eta j}^-]=0, \nn\\ 
&&[b_{-i}^+, b_{-j}^-]=[b_{+i}^-, b_{+j}^+], \qquad i\neq j, \nn\\
&&[b_{00}^+, b_{-j}^-]=[b_{00}^-, b_{+j}^+], \nn\\
&&[b_{00}^+, b_{+j}^-]=[b_{00}^-, b_{-j}^+], \label{B1R}\\
&&[[b_{\xi i}^+, b_{\eta j}^-], b_{\epsilon k}^+]=
\delta_{ij}\delta_{\xi \eta} b_{\epsilon k}^+ + 
\delta_{jk}\delta_{ \eta\epsilon } b_{\xi i}^+ - 
\delta_{ik}\delta_{\xi, -\epsilon} b_{-\eta j}^+,  \nn\\
&&[[b_{\xi i}^+, b_{\eta j}^-], b_{\epsilon k}^-]=
-\delta_{ij}\delta_{\xi \eta} b_{\epsilon k}^- - 
\delta_{ik}\delta_{ \xi\epsilon } b_{\eta j}^- + 
\delta_{jk}\delta_{\eta, -\epsilon} b_{-\xi i}^-. \nn
\end{eqnarray}

\smallskip \noindent {\bf Step 2.} 
Delete node $i$ ($i=2,\ldots,n$) from the Dynkin diagram; then the corresponding 
subalgebra is $G_0=H+sl(i)\oplus so(2(n-i)+1)$. Now there are four $G_0$-modules,
with the following grading for $G$:
\[
so(2n+1)=G_{-2}\oplus G_{-1}\oplus G_0\oplus G_{+1} \oplus G_{+2}
\]
with 
\begin{eqnarray}
G_{-1}&=&\hbox{span}\{ e_{j, 2n+1}-e_{2n+1,n+j},\  e_{j,k+n}-e_{k,n+j},\ 
 e_{jk}-e_{k+n,n+j}; \nn\\
 &&\ j=1,\ldots , i,\ k=i+1,\ldots ,n\},\label{GB2}\\
G_{-2}&=&\hbox{span}\{ e_{j,k+n}-e_{k,j+n}; \ 1\leq j<k\leq i\}. \nn
\end{eqnarray}
The number of the annihilation operators is $N=2i(n-i)+i$. The most interesting case is 
that with $i=n$: this is the para-Fermi case presented in the Introduction. Indeed, let
\begin{equation}
f_{j}^-= \sqrt{2}(e_{j, 2n+1}-e_{2n+1,n+j}), \quad 
 f_{j}^+=\sqrt{2}(e_{2n+1,j}-e_{n+j,2n+1}), \qquad j=1,\ldots , n. 
\end{equation}
Then there are no quadratic relations, and ${\cal R}$ consists of triple relations only:
\begin{eqnarray}
&& [[f_{ j}^{\xi}, f_{ k}^{\eta}], f_{l}^{\epsilon}]=\frac 1 2
(\epsilon -\eta)^2
\delta_{kl} f_{j}^{\xi} -\frac 1 2  (\epsilon -\xi)^2
\delta_{jl}f_{k}^{\eta},  \label{GB3}\\
&& \qquad\qquad \xi, \eta, \epsilon =\pm\hbox{ or }\pm 1;\quad j,k,l=1,\ldots,n. \nn 
\end{eqnarray}

\smallskip \noindent {\bf Step 3.} 
Delete two or more nodes from the Dynkin diagram. Then the corresponding $\Z$-grading
of $so(2n+1)$ has no longer the required properties (i.e.\ there are non-zero $G_i$ with $|i|>2$).

\smallskip \noindent {\bf Step 4.}
Now we turn to the extended Dynkin diagram.
Deleting node $i$ from this diagram, leaves the Dynkin diagram of $so(2n+1)$ for $i=0,1$, 
of $so(2n)$ for $i=n$, of $sl(2)\oplus sl(2)\oplus so(2n-3)$ for $i=2$, of $sl(4)\oplus so(2n-5)$ for
$i=3$, and of $so(2i)\oplus so(2n-2i+1)$ for $i\geq 4$. In all these cases there is only one 
$G_0$-module, so there are no contributions to our classification.

\smallskip \noindent {\bf Step 5.}
Delete the adjacent nodes $(i-1)$ and $i$ ($i=3,\ldots ,n$) from the extended 
Dynkin diagram. The remaining diagram is that of
$\tilde{G}_0=sl(2)\oplus sl(2) \oplus so(2(n-i)+1)$ for $i=3$, of 
$\tilde{G}_0=sl(4) \oplus so(2(n-i)+1)$ for $i=4$, and of $\tilde{G}_0=so(2(i-1))
\oplus so(2(n-i)+1)$ for $i>4$. In each case, there are five $\tilde{G}_0$-modules $g_k$,
one of which is invariant under $\omega$ (say $g_1$). Then one has to put
$G_0=H+\tilde{G}_0 + g_1$, and in each case one finds $G_0 \equiv H+B_{n-1}$.

Now, there are only two $G_0$-modules and
\[
so(2n+1)=G_{-1}\oplus G_0\oplus G_{+1}
\]
with 
\begin{equation}
G_{-1}=\hbox{span}\{ e_{i, 2n+1}-e_{2n+1,n+i},\  e_{ik}-e_{k+n,n+i},\ 
 e_{i,k+n}-e_{k,n+i};\ k\neq i=1,\ldots , n\} .
\end{equation}
The number of the anihilation operators is $N=2n-1$, and all these cases are isomorphic to
those of Step~1.

\smallskip \noindent {\bf Step 6.}
Delete two nonadjacent nodes from the extended Dynkin diagram, say 
$i$ and $j$, $i<j$, $i,j\neq 0,1$. The remaining diagram is that of
$\tilde{G}_0=so(2i)\oplus sl(j-i)\oplus so(2(n-j)+1)$ (if $i=2$ we have $sl(2)\oplus sl(2)$
instead of $so(2i)$). There are seven $\tilde{G}_0$-modules $g_k$,
one of which (say $g_1$) with $\omega(g_1)=g_1$. Thus one has to
take $G_0=H+\tilde{G}_0 + g_1$, and this is in fact $G_0 \equiv H+so(2(n-j+i)+1)\oplus sl(j-i)$

The corresponding grading is:
\[
so(2n+1)=G_{-2}\oplus G_{-1}\oplus G_0\oplus G_{+1} \oplus G_{+2}
\]
with
\begin{eqnarray}
G_{-1}&=& \hbox{span}\{ e_{k, 2n+1}-e_{2n+1,n+k},\  e_{kl}-e_{l+n,n+k},\ 
 e_{k,n+l}-e_{l,n+k};\nn\\
 && k=i+1,\ldots , j,\ l=1,\ldots ,i,j+1,\ldots ,n\},\nn\\
G_{-2}&=&\hbox{span}\{ e_{k,n+l}-e_{l,n+k}; \ i+1\leq k<l\leq j\} .
\end{eqnarray}
The number of the annihilation operators is $N=2(j-i)(n-j+i)+j-i$, and
all these cases turn out to be isomorphic to those of Step~2.

\smallskip \noindent {\bf Step 7.}
If we delete 3 or more nodes from the extended Dynkin diagram, the corresponding
$\Z$-grading of $so(2n+1)$ has no longer the required properties (i.e.\ there are
non-zero subspaces $G_i$ with $|i|>2$).

\setcounter{equation}{0}
\section{The Lie algebra $C_n=sp(2n)$} \label{sec:C}%

$G=sp(2n)$ is the subalgebra of $sl(2n)$ consisting of matrices of the form:
\begin{equation}
\left(\begin{array}{cc} a&b  \\
c&-a^t
\end{array}\right),
\label{sp(2n)}
\end{equation}
where $a$ is any $(n\times n)$-matrix, and $b$ and $c$ are symmetric $(n\times n)$-matrices.
The Cartan subalgebra $H$ consist of the diagonal matrices, and the root vectors and 
corresponding roots of $G$ are:
\begin{eqnarray*}
e_{jk}-e_{k+n,j+n} & \leftrightarrow & \epsilon_j -\epsilon_k, \qquad j\neq k=1,\ldots ,n,\\
e_{j,k+n}+e_{k,j+n} & \leftrightarrow & \epsilon_j +\epsilon_k, \qquad j\leq k=1,\ldots ,n,\\
e_{j+n,k}+e_{k+n,j} & \leftrightarrow & -\epsilon_j -\epsilon_k, \qquad j\leq k=1,\ldots ,n.
\end{eqnarray*}
The simple roots, Dynkin diagram and extended Dynkin diagram are given in Table~1.
Again, the anti-involution is such that $\omega(e_{jk})=e_{kj}$.
Next, we describe the process of deleting nodes and its consequences for
the classification of GQS.

\smallskip \noindent {\bf Step 1.}
Delete node $i$ ($i=1,\ldots,n-1$) from the Dynkin diagram. The remaining diagram 
is that of $sl(i)\oplus sp(2(n-i))$, so $G_0=H+sl(i)\oplus sp(2(n-i))$. There are four $G_0$-modules,
leading to the following grading:
\[
sp(2n)=G_{-2}\oplus G_{-1}\oplus G_0\oplus G_{+1} \oplus G_{+2}
\]
with 
\begin{eqnarray}
G_{-1}&=&\hbox{span}\{ e_{k, n+l}+e_{l,n+k},\  e_{kl}-e_{n+l,n+k};\ 
  k=1,\ldots , i,\ l=i+1,\ldots ,n\}, \label{GC3}\\
G_{-2}&=&\hbox{span}\{ e_{k,n+l}+e_{l,n+k}; \ 1\leq k\leq l\leq i\}.\nn
\end{eqnarray}
The number of the annihilation operators is $N=2i(n-i)$. The most interesting cases 
are $i=1$ and $i=n-1$, which we shall describe in more detail.

For $i=1$, let us denote the CAOs by
\begin{eqnarray}
&&c_{-j}^-=e_{1,n+j+1}+e_{j+1,n+1}, \ c_{+j}^-=e_{1,j+1}-e_{n+j+1,n+1},\quad j=1,\ldots ,n-1,\nn\\
&&c_{-j}^+=e_{n+j+1,1}+e_{n+1,j+1}, \ c_{+j}^+=e_{j+1,1}-e_{n+1,n+j+1},\quad j=1,\ldots ,n-1. \label{C1}
\end{eqnarray}
Then the corresponding relations ${\cal R}$ read, with $\xi, \eta, \epsilon, \gamma =\pm$ or $\pm 1$,
and $j,k,l=1,\ldots,n-1$:
\begin{eqnarray}
&&[c_{\xi j}^{\eta },c_{\xi k}^{\eta}]=0,\nn\\
&&[c_{-j}^+,c_{- k}^-]=[c_{+j}^-,c_{+k}^+], \qquad j\neq k, \nn \\
&&[c_{-j}^-,c_{+ k}^-]=[c_{-j}^+,c_{+k}^+]=0, \qquad j\neq k,\nn \\
&&[[c_{\xi j}^+,c_{\eta k}^-],c_{\epsilon l}^+]=\delta_{\xi\eta}\delta_{jk}
 c_{\epsilon l}^+ +\delta_{\eta\epsilon}\delta_{kl}c_{\xi j}^+ +
 (-1)^{\eta\epsilon}\delta_{\xi ,-\epsilon}\delta_{jl}c_{-\eta k}^+, \label{C1R}\\
&&[[c_{\xi j}^+,c_{\eta k}^-],c_{\epsilon l}^-]=-\delta_{\xi\eta}\delta_{jk}
 c_{\epsilon l}^- -\delta_{\xi\epsilon}\delta_{jl}c_{\eta k}^- +
 (-1)^{\xi\eta}\delta_{\eta, -\epsilon}\delta_{kl}c_{-\xi j}^-,\nn\\
&&[[c_{- j}^{\xi},c_{+ k}^{\xi}],c_{\eta l}^{-\xi}]=2\eta \delta_{jk}c_{-\eta l}^{\xi},\nn\\
&&[[c_{\xi j}^\gamma,c_{\eta k}^\gamma],c_{\epsilon l}^\gamma]=0.\nn
\end{eqnarray}

For $i=n-1$, let us also denote the CAOs by $c_j^\pm$:
\begin{eqnarray}
&&c_{-j}^-=e_{j,2n}+e_{n,n+j}, \ c_{+j}^-=e_{jn}-e_{2n,n+j},\quad j=1,\ldots ,n-1,\nn\\ 
&&c_{-j}^+=e_{2n,j}+e_{n+j,n}, \ c_{+j}^+=e_{nj}-e_{n+j,2n},\quad j=1,\ldots ,n-1. \label{C2}
\end{eqnarray}
Now, the corresponding relations read, with $\xi, \eta, \epsilon, \gamma =\pm$ 
or $\pm 1$, $j,k,l=1,\ldots,n-1$:
\begin{eqnarray}
&&[c_{\xi j}^{\eta },c_{\xi k}^{\eta}]=0,\nn\\
&& [c_{+j}^+,c_{- k}^-]=[c_{+j}^-,c_{-k}^+]=0, \qquad j\neq k, \nn \\
&&[[c_{\xi j}^{\epsilon},c_{\xi k}^{-\epsilon}],c_{\eta l}^{\epsilon}]=
 \xi \eta \delta_{jk} c_{\eta l}^{\epsilon}+\delta_{kl} c_{\eta j}^{\epsilon}, \nn\\
&&[[c_{+ j}^{\epsilon},c_{- k}^{-\epsilon}],c_{\eta l}^{\xi}]=
(\epsilon\xi-\eta)\delta_{jk} c_{-\eta l}^{\xi},\label{C2R}\\
&&[[c_{+ j}^{\epsilon},c_{- k}^{\epsilon}],c_{\xi l}^{-\epsilon}]=
-\xi \delta_{jl} c_{-\xi k}^{\epsilon}- \xi \delta_{kl}c_{-\xi j}^{\epsilon},\nn\\
&&[[c_{\xi j}^{\gamma},c_{\eta k}^{\gamma}],c_{\epsilon l}^{\gamma}]=0.\nn
\end{eqnarray}
This set of CAOs, together with their relations~(\ref{C2R}),  was constructed earlier 
in~\cite{Palev1}. Also the CAOs~(\ref{C1}) were already mentioned in~\cite{Palev1}
as a possible example, without giving the actual relations~(\ref{C1R}).

\smallskip \noindent {\bf Step 2.}
When node $n$ is deleted from the Dynkin diagram of $C_n$, the corresponding diagram is 
that of $sl(n)$, and $G_0=H+sl(n)$. In this case, there are two $G_0$-modules, and
$sp(2n)$ has the grading $sp(2n)=G_{-1}\oplus G_0\oplus G_{+1}$ with 
\begin{equation}
G_{-1}=\{ e_{j, n+k}+e_{k,n+j};\   
1\leq j\leq  k\leq n\}. \label{C3}
\end{equation}
There are $N=\frac{n(n+1)} {2}$ commuting annihilation operators, and
the relations ${\cal R}$ will not be given explicitly.

\smallskip \noindent {\bf Step 3.}
Upon deleting two or more nodes from the Dynkin diagram of $C_n$, the corresponding $\Z$-gradings
have no longer the required property (there are non-zero $G_i$ with $|i|>2$).

\smallskip \noindent {\bf Step 4.}
Now we turn to the extended Dynkin diagram. Deleting one node from this diagram
leads to a situation with only one $G_0$-module, irrelevant for our classification.

\smallskip \noindent {\bf Step 5.}
Delete the adjacent nodes $(i-1)$ and $i$ ($i=2,\ldots ,n$) from the extended 
Dynkin diagram. The remaining diagram is that of $\tilde{G}_0=sp(2(i-1))
\oplus sp(2(n-i))$. There are seven $\tilde{G}_0$-modules $g_k$, one of which
satisfies $\omega(g_1)=g_1$. Putting $G_0=H+\tilde{G}_0 + g_1$, it turns out
that $G_0 \equiv H+C_{n-1}$. In that case, there are only four $G_0$-modules and
$G$ has the grading $sp(2n)=G_{-2}\oplus G_{-1}\oplus G_0\oplus G_{+1} \oplus G_{+2}$
with 
\begin{equation}
G_{-1}=\hbox{span}\{ e_{i, n+j}+e_{j,n+i},\  e_{ij}-e_{n+j,n+i};\ 
  j\neq i=1,\ldots , n\}.
\end{equation}
The number of the annihilation operators is $N=2(n-1)$, and all these cases are isomorphic to
the $i=1$ case of Step~1.

\smallskip \noindent {\bf Step 6.}
Delete two nonadjacent nodes  $i<j$  (excluding the case $i=1$ and $j=n$)
from the extended Dynkin diagram. The remaining diagram is that of 
$\tilde{G}_0=sp(2i)\oplus sl(j-i)\oplus sp(2(n-j))$. There are again seven 
$\tilde{G}_0$-modules $g_k$, among which one with $\omega(g_1)=g_1$. 
Then $G_0=H+\tilde{G}_0 + g_1\equiv H+sl(j-i)\oplus sp(2(n-j+i))$. 
There are only four $G_0$-modules and the grading is 
$sp(2n)=G_{-2}\oplus G_{-1}\oplus G_0\oplus G_{+1} \oplus G_{+2}$
with 
\begin{equation}
G_{-1}=\hbox{span}\{ e_{k, n+l}+e_{l,n+k},\  e_{kl}-e_{n+l,n+k};\ 
k=i+1,\ldots , j,\  l\neq i+1,\ldots , j\}.
\end{equation}
The number of annihilation operators is $N=2(j-i)(n-j+i)$,
and all these cases are isomorphic to those of Step~1 with $i\neq 1$.

\smallskip \noindent {\bf Step 7.}
Delete node~1 and $n$ from the extended Dynkin diagram. The remaining
diagram is that of $sl(2)\oplus sl(n-1)$. With $G_0=sl(2)\oplus sl(n-1)$, 
there are four $G_0$-modules and the corresponding grading is 
$sp(2n)=G_{-2}\oplus G_{-1}\oplus G_0\oplus G_{+1} \oplus G_{+2}$
with 
\begin{equation}
G_{-1}=\hbox{span}\{ e_{1, n+k}+e_{k,n+1},\  e_{k1}-e_{n+1,n+k};\ 
  k=2,\ldots , n\}.
\end{equation}
This case is isomorphic to the $i=n-1$ case of Step~1.

\smallskip \noindent {\bf Step 8.}
If we delete 3 or more nodes from the extended Dynkin diagram, the corresponding
$\Z$-grading of $sp(2n)$ has no longer the required properties (i.e.\ there are
non-zero subspaces $G_i$ with $|i|>2$).

\setcounter{equation}{0}
\section{The Lie algebra $D_n=so(2n)$} \label{sec:D}%

$G=so(2n)$ is the subalgebra of $sl(2n)$ consisting of matrices of the form:
\begin{equation}
\left(\begin{array}{cc} a&b  \\
c&-a^t
\end{array}\right),
\label{so(2n)}
\end{equation}
where $a$ is any $(n\times n)$-matrix, and $b$ and $c$ are antisymmetric $(n\times n)$-matrices.
The Cartan subalgebra $H$ consist of the diagonal matrices, and the root vectors and 
corresponding roots of $G$ are:
\begin{eqnarray*}
e_{jk}-e_{k+n,j+n} & \leftrightarrow & \epsilon_j -\epsilon_k, \qquad j\neq k=1,\ldots ,n,\\
e_{j,k+n}-e_{k,j+n} & \leftrightarrow & \epsilon_j +\epsilon_k, \qquad j<k=1,\ldots ,n,\\
e_{j+n,k}-e_{k+n,j} & \leftrightarrow & -\epsilon_j -\epsilon_k, \qquad j<k=1,\ldots ,n.
\end{eqnarray*}
The simple roots, Dynkin diagram and extended Dynkin diagram are given in Table~1.
Again, the anti-involution is such that $\omega(e_{jk})=e_{kj}$.
Next, we describe the process of deleting nodes and its consequences for
the classification of GQS.

\smallskip \noindent {\bf Step 1.}
When node~1 is deleted from the Dynkin diagram of $D_n$, the remaining
diagram is that of $D_{n-1}$, so $G_0=H+D_{n-1}=H+so(2(n-1))$. There are two $G_0$-modules,
\begin{equation}
G_{-1}=\hbox{span}\{ e_{1i}-e_{n+i,n+1},\ e_{1,n+i}-e_{i,n+1};\ 
 i=2,\ldots ,n\}, \label{G1D4}
\end{equation}
and $G_{+1}=\omega(G_{-1})$. $G$ has the corresponding grading 
$so(2n)=G_{-1}\oplus G_0\oplus G_{+1}$, and there are $N=2(n-1)$ commuting 
annihilation operators.
Denoting the CAOs by
\begin{eqnarray}
&& d_{-i}^-=e_{1,n+i+1}-e_{i+1,n+1}, \ d_{+i}^-=e_{1,i+1}-e_{n+i+1,n+1},\qquad i=1,\ldots ,n-1,\nn\\
&& d_{-i}^+=e_{n+i+1,1}-e_{n+1,i+1}, \ d_{+i}^+=e_{i+1,1}-e_{n+1,n+i+1},\qquad i=1,\ldots ,n-1, \label{D1}
\end{eqnarray}
then, for $\xi, \eta, \epsilon =\pm$ and $ i,j,k=1,\ldots,n-1$, the relations
${\cal R}$ are given by:
\begin{eqnarray}
&&[d_{\xi i}^{\epsilon},d_{\eta j}^{\epsilon}]=0,\nn \\
&&[d_{- i}^{+},d_{+ i}^{-}]=[d_{+ i}^{+},d_{-i}^{-}]=0, \label{D1R} \\
&&[[d_{\xi i}^{+},d_{\eta j}^{-}],d_{\epsilon k}^{-}]=
-\delta_{\xi\eta} \delta_{ij}d_{\epsilon k}^{-}- \delta_{\xi\epsilon} \delta_{ik}
d_{\eta j}^{-}+\delta_{\eta,-\epsilon}\delta_{jk}d_{-\xi,i}^-, \nn\\
&&[[d_{\xi i}^{+},d_{\eta j}^{-}],d_{\epsilon k}^{+}]=
\delta_{\xi\eta} \delta_{ij}d_{\epsilon k}^{+}+ \delta_{\eta\epsilon} \delta_{jk}
d_{\xi i}^{+}-\delta_{\xi,-\epsilon}\delta_{ik}d_{-\eta,j}^+ .\nn
\end{eqnarray}
Although the relations~(\ref{D1R}) are new, the existence of the set of CAOs~(\ref{D1})
was pointed out in~\cite{Palev1}.

\smallskip \noindent {\bf Step 2.}
When node $i$ ($i=2,\ldots,n-2$) is deleted from the Dynkin diagram of $D_n$, the 
remaining diagram is that of $sl(i)\oplus so(2(n-i))$ 
(or $sl(n-2)\oplus sl(2)\oplus sl(2)$ in the case $i=n-2$). 
With $G_0=sl(i)\oplus so(2(n-i))$, there are four $G_0$-modules, and
$so(2n)$ has the following grading 
$so(2n)=G_{-2}\oplus G_{-1}\oplus G_0\oplus G_{+1} \oplus G_{+2}$
with 
\begin{equation}
G_{-1}=\hbox{span}\{ e_{kl}-e_{n+l,n+k},\  e_{k,n+l}-e_{l,n+k};\ 
k=1,\ldots , i,\ l=i+1,\ldots ,n\}. \label{G2D4}
\end{equation}
The number of annihilation operators is $N=2i(n-i)$. 

\smallskip \noindent {\bf Step 3.}
Delete node $n-1$ or $n$ from the Dynkin diagram; the remaining diagram
is that of $sl(n)$, and $G_0=H+sl(n)$. There are only two $G_0$-modules and
$G$ has the grading $so(2n)=G_{-1}\oplus G_0\oplus G_{+1}$, with 
\begin{eqnarray}
G_{-1}&=&\hbox{span} \{ e_{j,n+k}-e_{k,n+j};\ 1\leq j<k\leq n-1\} \cup \nn\\
&& \hbox{span}\{e_{jn}-e_{2n,n+j};\ j=1,\ldots , n-1\},\hbox{ for } i=n-1, \label{G3D4}\\
G_{-1}&=&\hbox{span}\{ e_{j,k+n}-e_{k,j+n}; \ 1\leq j<k\leq n\}, \hbox{ for } i=n.\nn
\end{eqnarray}
There are $N=\frac{n(n-1)}{2}$ commuting annihilation operators, and
these two cases are isomorphic. The relations are not given explicitly.

\smallskip \noindent {\bf Step 4.}
Upon deleting two nodes $i$ and $j$ ($i<j=1,\ldots, n-2$) or more from the 
Dynkin diagram of $D_n$, the corresponding $\Z$-gradings
have no longer the required property (there are non-zero $G_i$ with $|i|>2$).

\smallskip \noindent {\bf Step 5.}
Delete nodes $n-1$ and $n$ from the Dynkin diagram. The remaining diagram is that of
$sl(n-1)$. For $G_0=H+sl(n-1)$, there are six $G_0$-modules. 
There are three different ways in which these $G_0$-modules can be combined,
each of them yielding a $\Z$-grading of the form
$so(2n)=G_{-2}\oplus G_{-1}\oplus G_0\oplus G_{+1} \oplus G_{+2}$, namely:
\begin{eqnarray}
G_{-1}&=&\hbox{span} \{ e_{jn}-e_{2n,n+j}, \  e_{j,2n}-e_{n,n+j};\ 
j=1,\ldots , n-1\}, \label{D2}\\
G_{-1}&=&\hbox{span}\{ e_{jn}-e_{2n,n+j}, \ j=1,\ldots , n-1;\nn\\
&& e_{n+j,k}-e_{n+k,j},\ 1\leq j<k\leq n-1\}, \label{D3}\\
G_{-1}&=&\hbox{span}\{ e_{j+n,n}-e_{2n,j}, \ j=1,\ldots , n-1;\nn\\
&& e_{j,k+n}-e_{k,j+n},\ 1\leq j<k\leq  n-1\}. \label{D4}
\end{eqnarray}
For (\ref{D2}), we have $N=2(n-1)$; for (\ref{D3}) and (\ref{D4}), we have $N=\frac{n(n-1)}{2}$.
It turns out that (\ref{D3}) and (\ref{D4}) are isomorphic to each other.
Here, we shall give the relations only for (\ref{D2}).
Denote the CAOs of (\ref{D2}) by
\begin{eqnarray}
&& d_{-i}^-=e_{i,2n}-e_{n,n+i}, \ d_{+i}^-=e_{in}-e_{2n,n+i},\qquad i=1,\ldots ,n-1,\nn\\
&& d_{-i}^+=e_{2n,i}-e_{n+i,n}, \ d_{+i}^+=e_{ni}-e_{n+i,2n},\qquad i=1,\ldots ,n-1. \label{D2cao}
\end{eqnarray}
Then, with $\xi, \eta, \epsilon, \gamma =\pm$ or $\pm 1$ and $i,j,k=1,\ldots,n-1$, the relations
are explicitly given by:
\begin{eqnarray}
&&[d_{\xi i}^{\eta},d_{\xi j}^{\eta}]=0,\nn \\
&&[d_{- i}^{+},d_{+ j}^{-}]=[d_{+ i}^{+},d_{-j}^{-}]=0,\nn\\
&&[d_{+ i}^{-},d_{- i}^{-}]=[d_{+ i}^{+},d_{-i}^{+}]=0,\nn\\
&&[[d_{\xi i}^{\gamma},d_{\eta j}^{\gamma}],d_{\epsilon k}^{\gamma}]=0,\label{D2R}\\
&&[[d_{+ i}^{\xi},d_{- j}^{\xi}],d_{\epsilon k}^{-\xi}]=
-\delta_{ik}d_{-\epsilon j}^{\xi}+  \delta_{jk} d_{-\epsilon i}^{\xi},\nn\\
&&[[d_{\xi i}^{\eta},d_{\xi j}^{-\eta}],d_{\epsilon k}^{\eta}]=
\xi \epsilon \delta_{ij}d_{\epsilon k}^{\eta}+  \delta_{jk} d_{\epsilon i}^{\eta}. \nn
\end{eqnarray}
The set of CAOs~(\ref{D2cao}) with relations~(\ref{D2R}) is the example that was considered
earlier in~\cite{Palev1} and \cite{Palev3}.

\smallskip \noindent {\bf Step 6.}
Now we move to the extended Dynkin diagram. Deleting node~$i$ leaves the Dynkin diagram
of $so(2n)$ for $i=0,1,n-1,n$, of $sl(2)\oplus sl(2)\oplus so(2(n-2))$ for $i=2$, of 
$sl(3)\oplus so(2(n-3))$ for $i=3$, and of $=so(2i)\oplus so(2(n-i))$ for $i\geq 4$. 
In all these cases there is only one $G_0$-module, so there are no contributions to
our classification.

\smallskip\noindent 
Note that deleting nodes $i$ and $j$ ($1<i<j<\lfloor\frac{n+1}{2}\rfloor$) 
from the extended Dynkin diagram is equivalent to delete nodes $(n-j)$ and $(n-i)$. 

\smallskip \noindent {\bf Step 7.}
Delete the adjacent nodes $(j-1)$ and $j$.
For $j=1$ we are back to Step~1, and for $j=2$ to Step~2 with $i=2$.
For $j\geq 3$ the remaining diagram is that of $\tilde{G}_0=so(2(j-1))\oplus so(2(n-j))$ 
(for $j=3$ this is $sl(2)\oplus sl(2)\oplus so(2(n-j))$ and for 
$j=4$ this is $sl(4)\oplus so(2(n-j))$). There are five $\tilde{G}_0$-modules
$g_k$, one with $\omega(g_5)=g_5$, so one has to put 
$G_0=H+\tilde{G}_0 + g_5\equiv H+  so(2(n-1))$. Now, there are only two $G_0$-modules,
$G$ has the grading $so(2n)= G_{-1}\oplus G_0\oplus G_{+1}$, and all 
these cases are isomorphic to those of Step~1.

\smallskip \noindent {\bf Step 8.}
Delete the nonadjacent nodes $i$ and $j$ ($i<j-1$) from the 
extended Dynkin diagram. The remaining diagram is that of
$\tilde{G}_0=so(2i)\oplus sl(j-i)\oplus so(2(n-j))$ (for $i=2$  this is
$sl(2)\oplus sl(2)\oplus sl(j-i)\oplus so(2(n-j))$; for $i=3$ this is
$sl(3)\oplus sl(j-i)\oplus so(2(n-j))$).
There are nine $\tilde{G}_0$-modules $g_k$, one with $\omega(g_9)=g_9$.
Putting $G_0=H+\tilde{G}_0 + g_9\equiv H+ sl(j-i)\oplus so(2(n-j+i))$,
there are only four $G_0$-modules. All these cases are isomorphic to those of Step~2.

\smallskip \noindent {\bf Step 9.}
If we delete 3 or more nodes from the extended Dynkin diagram, the corresponding
$\Z$-grading of $so(2n)$ has no longer the required properties (i.e.\ there are
non-zero subspaces $G_i$ with $|i|>2$).

\setcounter{equation}{0}
\section{Summary and conclusions} \label{sec:concl}%

We have obtained a complete classification of all GQS associated with 
the classical Lie algebras. The familiar cases (para-Fermi statistics and
$A$-statistics) appear as simple examples in our classification. It is
worth observing that some other examples in this classification are
also rather simple. The GQS given in (\ref{A2R}) and (\ref{A21R}), e.g., 
seem to be closely related to $A$-statistics, 
except that there are two kind of `particles' corresponding to
the CAOs (see (\ref{A2CAO}) and (\ref{twokinds})). 
The GQS of type $D$ given in (\ref{D1R})
has also particularly simple defining relations. For convenience, 
a comprehensive summary of the classification of all GQS is given in Table 2.

As we have already mentioned in the main text, several cases in our classification
appear as examples in Ref.~\cite{Palev2}-\cite{Jellal} and in  
Palev's thesis~\cite{Palev1}. In these papers or in the thesis, however, no 
classification is given: only a number of examples inspired by the para-Fermi case
are considered. Furthermore, for some of these examples Fock type representations
are constructed.

In order to study the physical properties of a GQS, one should determine
the action of the CAOs in a Fock space. Thus one is automatically led
to representation theory. Here, the Lie algebraic framework is useful,
since a lot is known about Lie algebra representations. 
Apart from other properties to be satisfied, these Fock spaces should be `unitary'
(with respect to the given anti-involution $\omega$).
Whether the class of finite dimensional representations of $G$ plays 
a role, or whether it is a class of infinite dimensional representations,
depends on the choice of $\omega$. With the standard choice considered in
this paper, the unitary representations are finite dimensional. For another
choice of $\omega$ (still with $\omega(G_{-1})=G_{+1}$, but no longer all
$+$-signs in $\omega(x_i^-)=\pm x_i^+$), our classification
of GQS remains valid, but the unitary representations will be infinite dimensional.

It is only after a classification of the Fock spaces for a particular
GQS that one can study its macroscopic and microscopic properties.
Such a program is feasable, and can give rise to interesting quantum
statistical properties. For example, for $A$-statistics, the microscopic
properties (i.e.\ the properties of the CAOs and their action on
the Fock spaces) have been described in~\cite{Palev2}-\cite{PalevJeugt}, 
whereas the macroscopic
properties (i.e.\ the statistical properties of ensembles of `particles'
satisfying this GQS) have been studied in~\cite{Jellal}.
We hope that some other cases of this classification will yield similar
interesting GQS.

From the mathematical point of view, a set of CAOs together with a complete set
of relations ${\cal R}$ unambiguously describes the Lie algebra. So each case
of our classification also gives the description of a classical Lie algebra
in terms of a number of generators subject to certain relations.
This can also be reformulated in terms of the notion of Lie triple 
systems~\cite{Jacobson}.
According to the definition, a Lie triple system $L$ of an associative
algebra $A$ is a subspace of $A$ that is closed under the ternary composition
$[[a,b],c]$, where $[a,b]=ab-ba$. 
It is easy to see that in our case the subspace $G_{-1}\oplus G_{+1}$
(i.e.\ the subspace spanned by all CAOs) is a Lie triple system for
the universal enveloping algebra $U(G)$.

This paper was devoted to classical Lie algebras only. The exceptional 
Lie algebras are not considered here. Although it would be possible 
to perform a mathematical classification of the GQS associated with 
$G_2$, $F_4$, $E_6$, $E_7$ and $E_8$, it is obvious that in such a case
the number of CAOs is a fixed integer. For physical applications, it is
of importance that the number of CAOs is not a fixed number but an integer 
parameter $N$. In fact, in quantum field theoretical applications, one is mainly
interested in the case $N\rightarrow \infty$.

As mentioned in the introduction, para-Bose statistics is connected
with a Lie superalgebra, the orthosymplectic superalgebra $osp(1|2n)$. 
In a future paper, we hope to classify all GQS associated with
the classical Lie superalgebras.

\bigskip
\noindent{\bf Acknowledgements}
\medskip

\noindent
The authors are thankful to Professor T.D. Palev for constructive discussions.
NIS was supported by a Marie Curie Individual Fellowship of
the European Community Programme `Improving the Human Research Potential and the
Socio-Economic Knowledge Base' under contract number HPMF-CT-2002-01571.

\newpage
\noindent
{\bf Table 1}.
Classical Lie algebras, their (extended) Dynkin diagrams with a labelling of
the nodes and the corresponding simple roots.
\vskip 1cm
\noindent
\begin{tabular}{ccc}
\hline
Lie algebra & Dynkin diagram & extended Dynkin diagram \\
\hline \\[5mm]
$A_n$ & & \\
$(n>0)$ & & \\[-15mm]
  & \mbox{\includegraphics{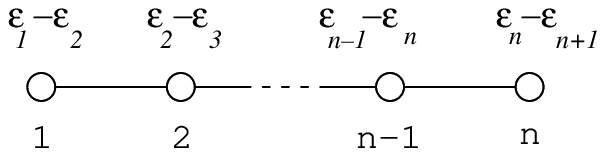}} & \includegraphics{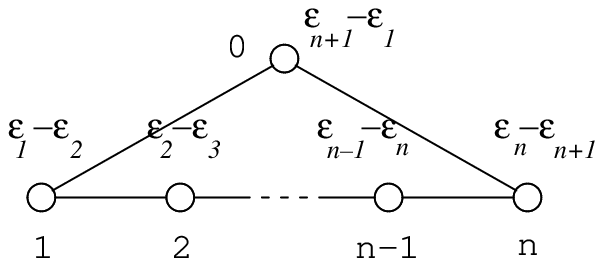}\\[10mm]
$B_n$ & & \\
$(n>1)$ & & \\[-15mm]
  & \mbox{\includegraphics{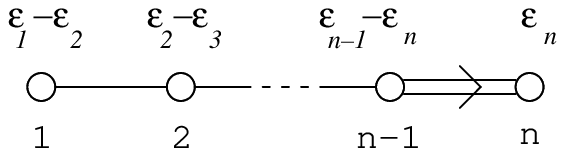}} & \includegraphics{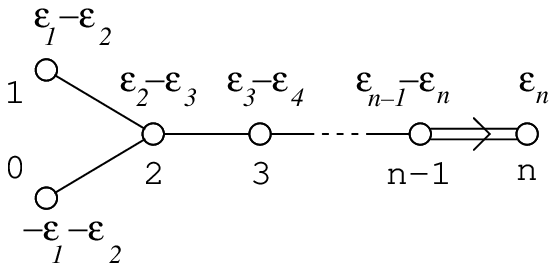}\\[10mm]
$C_n$ & & \\
$(n>2)$ & & \\[-10mm]
  & \mbox{\includegraphics{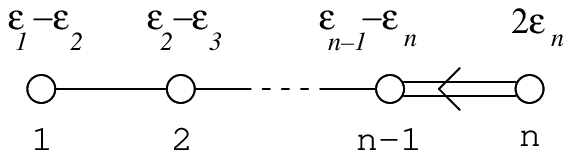}} & \includegraphics{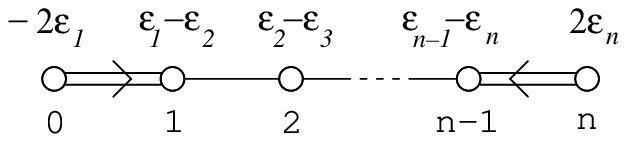}\\[10mm]
$D_n$ & & \\
$(n>3)$ & & \\[-15mm]
  & \mbox{\includegraphics{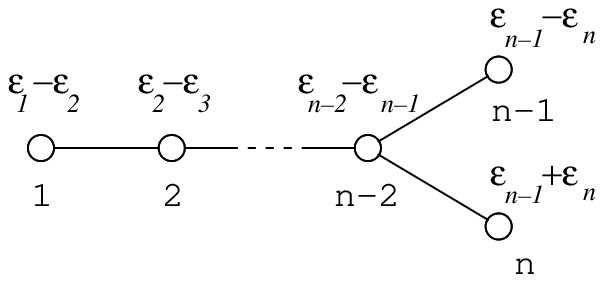}} & \includegraphics{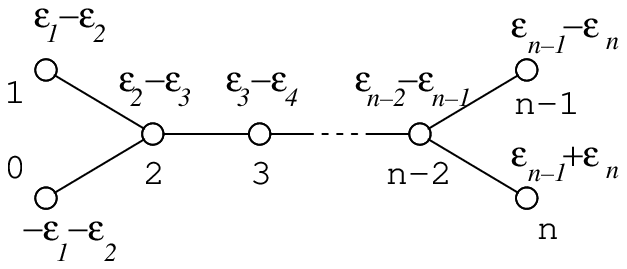}\\
\hline
\end{tabular}

\newpage
\noindent
{\bf Table 2}.
Summary of the classification: all non-isomorphic GQS associated with a 
classical Lie algebra are given. For each GQS, we list: the Dynkin diagram
of $G_0$ (described in terms of the  Dynkin diagram $D$ 
of $G$), the subspace $G_{-1}$ (as a reference to the main text), 
the number of annihilation operators ($N$), and
the relations ${\cal R}$ (if given in the text).
\vskip 1cm
\noindent
{\small
\begin{tabular}{l|l|l|l|l}
\hline
Lie     & Dynkin diagram & $G_{-1}$ & $N$ & ${\cal R}$ \\
algebra & of $G_0$       &          &     &            \\ 
\hline \hline
$A_n$ 
  & $\begin{array}{l} D - \{i\} \\ (i\leq \lfloor \frac{n+1}{2}\rfloor) \end{array}$ 
  & (\ref{A1G-1}) 
  & $i(n+1-i)$ 
  & $\begin{array}{l} i=1:\ (\ref{A1R}) \\ i=2:\ (\ref{A2R}) \end{array}$ 
  \\
\cline{2-5}
  & $\begin{array}{l} D - \{i,j\} \\ (i\leq \lfloor \frac{n}{2}\rfloor\\ i<j<n+1-i) \end{array}$ 
  & (\ref{A21}) 
  & $(j-i)(n+1-j+i)$ 
  & $j-i=1:\ (\ref{A21R})$ 
  \\
\hline  
$B_n$ 
  & $D - \{1\}$ 
  & (\ref{GB1}) 
  & $2n-1$ 
  & (\ref{B1R}) 
  \\
\cline{2-5}
  & $\begin{array}{l} D - \{i\} \\ (2\leq i\leq n) \end{array}$ 
  & (\ref{GB2}) 
  & $2i(n-i)+i$ 
  & $i=n$: (\ref{GB3})  
  \\
\hline 
$C_n$ 
  & $\begin{array}{l} D - \{i\} \\ (1\leq i\leq n-1) \end{array}$ 
  & (\ref{GC3}) 
  & $2i(n-i)$ 
  & $\begin{array}{l} i=1:\ (\ref{C1R}) \\ i=n-1:\ (\ref{C2R}) \end{array}$ 
  \\
\cline{2-5}
  & $D - \{n\}$ 
  & (\ref{C3}) 
  & ${\frac{n(n+1)}{2}}$ 
  &  -
  \\ 
\hline  
$D_n$ 
& $D - \{1\}$ 
  & (\ref{G1D4}) 
  & $2(n-1)$ 
  &  (\ref{D1R}) 
  \\ 
 \cline{2-5} 
  & $\begin{array}{l} D - \{i\} \\ (2\leq i\leq n-2) \end{array}$ 
  & (\ref{G2D4}) 
  & $2i(n-i)$ 
  & -
  \\
\cline{2-5}
  & $D - \{n\}$ 
  & (\ref{G3D4}) 
  & ${\frac{n(n-1)}{2}}$ 
  &  -
  \\ 
\cline{2-5}
 &$D - \{n-1,n\}$
 & $\begin{array}{l}(\ref{D2}) \\ (\ref{D3}) \end{array}$ 
 & $\begin{array}{l} 2(n-1) \\ {\frac{n(n-1)}{2}} \end{array}$     
 & $\begin{array}{l}(\ref{D2R}) \\ $-$ \end{array}$ 
 \\ 
\hline  
\end{tabular}
}

\end{document}